\documentclass[floats, aps, showpacs, twocolumn, superscriptaddress, prl,
    reprint, floatfix]{revtex4-1}

\usepackage{graphicx}
\usepackage{amsmath, amssymb, amstext, amsthm, amsfonts}
\usepackage{bbm}

\usepackage{times}

\begin{document}

\title{Localization of Chaotic Resonance States due to a Partial Transport
Barrier}

\author{Martin J.\ K\"orber}
\affiliation{Technische Universit\"at Dresden,
             Institut f\"ur Theoretische Physik and Center for Dynamics,
             01062 Dresden, Germany}

\author{Arnd B\"acker}
\affiliation{Technische Universit\"at Dresden,
             Institut f\"ur Theoretische Physik and Center for Dynamics,
             01062 Dresden, Germany}
\affiliation{Max-Planck-Institut f\"ur Physik komplexer Systeme, N\"othnitzer
             Stra\ss{}e 38, 01187 Dresden, Germany}

\author{Roland Ketzmerick}
\affiliation{Technische Universit\"at Dresden,
             Institut f\"ur Theoretische Physik and Center for Dynamics,
             01062 Dresden, Germany}
\affiliation{Max-Planck-Institut f\"ur Physik komplexer Systeme, N\"othnitzer
             Stra\ss{}e 38, 01187 Dresden, Germany}

\date{\today}

\begin{abstract}
Chaotic eigenstates of quantum systems are known to localize on either side of
a classical partial transport barrier if the flux connecting the two sides is
quantum mechanically not resolved due to Heisenberg's uncertainty.
Surprisingly, in open systems with escape chaotic resonance states can localize
even if the flux is quantum mechanically resolved. We explain this using the
concept of conditionally invariant measures from classical dynamical systems by
introducing a new quantum mechanically relevant class of such fractal measures.
We numerically find quantum-to-classical correspondence for localization
transitions depending on the openness of the system and on the decay rate of
resonance states.
\end{abstract}

\maketitle

%%%%%%%%%%%%%%%%%%%%%%%%%%%%%%%%%%%%%%%%%%%%%%%%%%%%%%%%%%%%%%%%%%%%%%%%%%%%%%%

Localization of quantum eigenstates and wave packets is of fundamental
importance for the physics of transport and appears for a variety of reasons,
e.g., strong localization due to disorder~\cite{And1958}, weak localization due
to time-reversal symmetry~\cite{Ber1984}, localized edge states due to
topological protection~\cite{QiZha2011}, or localization due to classically
restrictive phase-space structures~\cite{BohTomUll1993}. In the latter case,
the localization can originate from impenetrable barriers of regular motion or
partial transport barriers with a small transmission given by a flux $\Phi$
within a chaotic region~\cite{KayMeiPer1984a, KayMeiPer1984b, BroWya1986,
GeiRadRub1986, Mei1992, BohTomUll1993, KetHufSteWei2000, MaiHel2000}. Such
partial barriers are ubiquitous in the chaotic region of generic two
degree-of-freedom Hamiltonian systems~\cite{KayMeiPer1984a, KayMeiPer1984b,
Mei1992} and a universal localization transition was
found~\cite{MicBaeKetStoTom2012}. Chaotic eigenstates of the system typically
localize on either side of a partial barrier if the transmission region is
quantum mechanically not resolved, i.e., if the classical flux $\Phi$ across
the partial barrier is much smaller than the size $h$ of Planck's cell~($\Phi
\ll h$). If the transmission region is quantum mechanically resolved~($h \ll
\Phi$), eigenstates are equipartitioned in the chaotic component, thereby
ignoring the presence of the partial barrier.

\begin{figure}
\begin{center}
\includegraphics[scale=0.95]{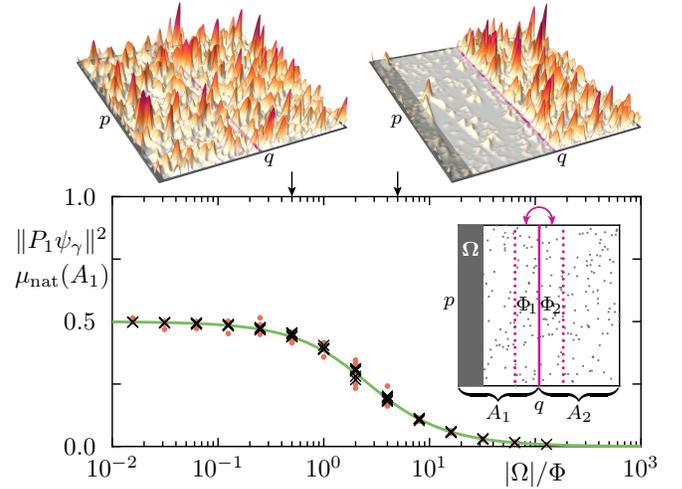}
\caption{
(color online) Weight $\|P_1\psi_\gamma\|^2$ (symbols) of resonance states in
region $A_1$ vs ratio of size $|\Omega|$ of opening and flux $\Phi$ across a
partial barrier for different parameters of the partial-barrier standard map
($16 \leq \Phi/h, |\Omega|/h \leq 2048$; $|A_1|=0.5$; $h=1/6000$). Weight of
state with $\gamma$ closest to $\gamma_\text{nat}$ (red points) and averaged
over states with decay rates $\gamma \in [\gamma_\text{nat}/1.1, 1.1
\,\gamma_\text{nat}]$ (black crosses). This is compared to the natural CIM
$\mu_\text{nat}(A_1)$ [Eq.~\eqref{eq:loc_classics_explicit}, solid green line].
Inset: Phase space of the partial-barrier map, illustrating regions $A_1$,
$A_2$ on either side of the partial barrier (solid magenta line) with
exchanging regions $\Phi_1$, $\Phi_2$, and opening $\Omega$.
Upper panels: Husimi representation of typical resonance states with $\gamma
\approx \gamma_\text{nat}$ for $h=1/1000$, $\Phi/h = 20$, and two values
$|\Omega|/\Phi$ indicated by arrows.
} \label{fig:fig1}
\end{center}
\end{figure}

In contrast, in open Hamiltonian systems which allow for
escape~\cite{AltPorTel2013, Nov2013, CasMasShe1999b, LuSriZwo2003, SchTwo2004,
KeaNovPraSie2006, NonSch2008, NonZwo2009, ErmCarSar2009, WeiBarKuhPolSch2014,
SchAlt2015}, chaotic resonance states exhibit localization in the presence of a
partial barrier surprisingly even in the semiclassical regime ($h \ll
\Phi$)~\cite{KoeMicBaeKet2013}. Such a localized state is shown in
Fig.~\ref{fig:fig1}, upper right, by its Husimi phase-space representation.
This demonstrates that in open systems the influence of partial barriers on
localization and transport properties is even more substantial than in closed
systems.
A thorough understanding of this localization phenomenon remains open, so far.
A prominent application are optical microcavities, where the emission patterns
are governed by the localization of eigenmodes~\cite{NoeSto1997,
GmaCapNarNoeStoFaiSivCho1998, LeeRimRyuKwoChoKim2004, WieMai2008, WieHen2008,
ShiLeeKimLeeYanMooLeeAn2008, ShiHarFukHenSasNar2010, ShiWieCao2011,
CaoWie2015}. For their design, it is particularly important to know whether a
partial barrier is desired to enhance localization or whether it should be
avoided. The localization phenomenon may also have relevance in many other
areas of physics, such as transport through quantum dots~\cite{Ihn2009},
ionization of driven Rydberg atoms~\cite{BucDelZak2002}, and microwave
cavities~\cite{Sto2007b}.

\begin{figure}
\begin{center}
\includegraphics[scale=0.97]{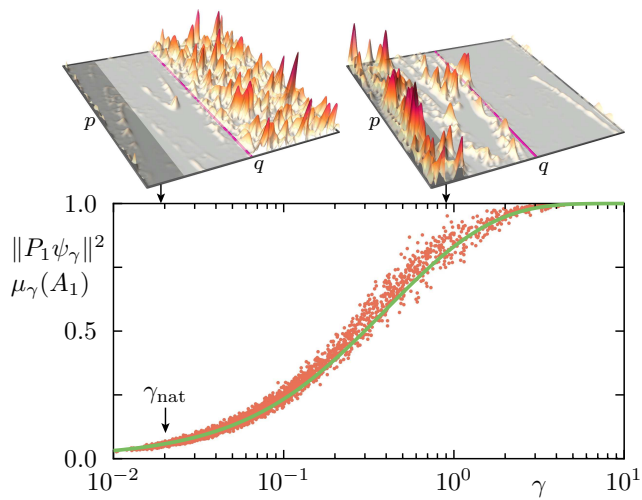}
\caption{
(color online) Weight $\|P_1\psi_\gamma\|^2$ (red points) of resonance states
$\psi_\gamma$ in region $A_1$ vs decay rate $\gamma$ for the partial-barrier
standard map ($\Phi/h = 64$; $|\Omega|/h = 1024$; $|A_1|=0.5$; $h=1/6000$).
This is compared to the $\gamma$-natural CIM $\mu_\gamma(A_1)$
[Eq.~\eqref{eq:loc_classics_explicit}, solid green line].
Upper panels: Husimi representation of typical long-lived (left) and
short-lived (right) resonance state for $h=1/1000$ with $\gamma$ values
indicated by arrows.
} \label{fig:fig2}
\end{center}
\end{figure}

Since the localization appears in a semiclassical regime ($h\ll\Phi$), one may
wonder if it has a classical origin. Thus one needs the classical
counterpart of a quantum resonance state. This is given in the field of open
dynamical systems~\cite{PiaYor1979, KanGra1985, Tel1987, DemYou2006,
NonRub2007, LaiTel2011, AltPorTel2013, AltPorTel2013b, MotGruKarTel2013} by a
conditionally invariant measure (CIM). It is invariant under time evolution up
to an exponential decay with rate $\gamma$. The asymptotic decay of generic
initial phase-space distributions leads to the so-called \emph{natural
CIM}~$\mu_\text{nat}$ with decay rate $\gamma_\text{nat}$. The
quantum-mechanical relevance of $\mu_\text{nat}$ is shown in
\cite{CasMasShe1999b, LeeRimRyuKwoChoKim2004, NonRub2007, Nov2013,
AltPorTel2013}. Note that the steady probability distribution introduced in the
context of optical microcavities~\cite{LeeRimRyuKwoChoKim2004} corresponds to
$\mu_\text{nat}$. The natural CIM~$\mu_\text{nat}$ for the single decay rate
$\gamma_\text{nat}$, however, cannot be the classical counterpart for all
quantum resonance states as they have a wide range of decay rates (see, e.g.,
Fig.~\ref{fig:fig2}). Exceptional CIMs with decay rate $\gamma$
different from $\gamma_\text{nat}$ have been discussed~\cite{DemYou2006,
NonRub2007}. In fact, for each $\gamma$ one can construct infinitely many CIMs.
It is an open question which of these CIMs correspond to quantum resonance
states for arbitrary $\gamma$. To answer this question one has to go beyond the
important results of Ref.~\cite{KeaNovPraSie2006} which relate the
total weight of a resonance state on each forward escaping set to its decay
rate.

In this Letter, we introduce the quantum mechanically relevant class of CIMs.
Their localization explains the localization of chaotic resonance states in the
presence of a partial barrier. In particular, we find~(i) a transition from
equipartition to localization when opening the system, Fig.~\ref{fig:fig1},
and~(ii) a transition from localization on one side of the partial barrier to
localization on the other side for resonance states with increasing decay rate,
Fig.~\ref{fig:fig2}. We numerically demonstrate quantum-to-classical
correspondence for a designed partial-barrier map and the generic standard map.

%%%%%%%%%%%%%%%%%%%%%%%%%%%%%%%%%%%%%%%%%%%%%%%%%%%%%%%%%%%%%%%%%%%%%%%%%%%%%%%

\emph{Partial-barrier map.}---We design a chaotic model map with a single
partial barrier (similar to Ref.~\cite{KoeMicBaeKet2013}), which allows for
numerically varying the flux across the partial barrier and for deriving the
classical localization, Eq.~\eqref{eq:loc_classics_explicit}. The
partial-barrier map $T = M\circ E\circ O$ is a composition of three maps: the
map $M$ describes the unconnected chaotic dynamics within two regions, $A_k$.
They decompose the phase space $\Gamma = [0,1) \times [0,1)$ into $A_1 = [0,
|A_1|) \times [0, 1)$ and its complement $A_2 = \Gamma \setminus A_1$; see the
inset in Fig.~\ref{fig:fig1}, where $|A_1|$ denotes the area of $A_1$ and by
normalization, $|A_2| = 1 - |A_1|$. The map $E$ induces a flux $\Phi$ between
$A_1$ and $A_2$ by exchanging regions $\Phi_k \subset A_k$ with $|\Phi_k| =
\Phi$. The map $O$ opens the system by the absorbing region $\Omega$, which is
contained in region $A_1$.

\begin{figure}
\begin{center}
\includegraphics{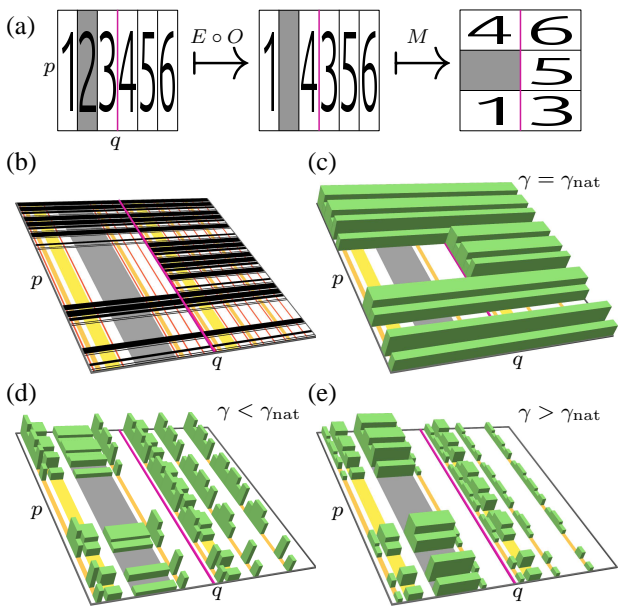}
\caption{
(color online) (a) Illustration of the partial-barrier Baker map $T=M\circ E
\circ O$. Magenta line indicates partial barrier and gray shaded region marks
the opening (left and central) and image of opening (right).
(b) Backward trapped set (dark horizontal stripes) and forward escaping sets
$\Omega$ (gray), $T^{-1}(\Omega)$ (yellow), $T^{-2}(\Omega)$ (orange), and
$T^{-3}(\Omega)$ (red).
(c) Natural CIM integrated over boxes of size $3^{-3}$ in the $p$ direction.
(d), (e) Approximation of $\gamma$-natural CIMs by truncation of
Eq.~\eqref{eq:cmeasure_arbitrary} to $n\leq 2$ for $\gamma \neq
\gamma_\text{nat}$.
} \label{fig:fig3}
\end{center}
\end{figure}

We introduce two different dynamics for $M$. For the numerical analysis, we use
the generic standard map~\cite{CasChiIzrFor1979} on the torus in symmetrized
form, $q_{t+1} = q_t + p_t^*$, $p_{t+1} = p_t^* + v(q_{t+1})$ with $p_t^* =
p_{t} + v(q_{t})$ for $v(q) = \frac{\kappa}{4\pi} \sin(2\pi q)$ acting
individually on each of the regions $A_k$ after appropriate rescaling. We fix
$\kappa = 10$ where the standard map displays a fully chaotic phase space. For
analytical considerations, we use the ternary Baker map in each region $A_k$,
as illustrated in Fig.~\ref{fig:fig3}(a), which allows for the derivation of
Eq.~\eqref{eq:loc_classics_explicit}. We refer to the corresponding maps $T$ as
\emph{partial-barrier standard map} and \emph{partial-barrier Baker map},
respectively.

%%%%%%%%%%%%%%%%%%%%%%%%%%%%%%%%%%%%%%%%%%%%%%%%%%%%%%%%%%%%%%%%%%%%%%%%%%%%%%%

\emph{Quantum localization transitions.}---Let us consider the quantization $U$
of the partial-barrier standard map $T$. From the eigenvalue problem for $U$,
\begin{equation}
U \psi_\gamma = e^{-\gamma/2}e^{i\theta} \psi_\gamma,
\end{equation}
we numerically compute the decay rates $\gamma$, describing the temporal decay
of the norm, $\|U^t \psi_\gamma\|^2 = e^{-\gamma t}$, and the corresponding
resonance states $\psi_\gamma$ (the phase $\theta$ is not relevant in the
following). The absolute weight of $\psi_\gamma$ in region $A_1$ is given by
$\|P_1 \psi_\gamma\|^2$, where $P_1$ denotes the projection onto the subspace
associated to $A_1$.
We observe (i) a transition from equipartition to localization on $A_2$ for
increasing size $|\Omega|$ of the opening, see Fig.~\ref{fig:fig1}, and (ii) a
transition from localization on $A_2$ to localization on $A_1$ for increasing
$\gamma$, see Fig.~\ref{fig:fig2}. Transition~(i) is surprising as localization
occurs for $h \ll \Phi$, where in the closed system all eigenstates are
equipartitioned~\cite{MicBaeKetStoTom2012}.
Transition~(ii) shows that in open systems the localization
depends on the decay rate $\gamma$.

In Fig.~\ref{fig:fig1}, we focus on resonances with decay rate $\gamma \approx
\gamma_\text{nat}$, which describes the decay of typical long-lived resonance
states in the semiclassical limit. We find transition~(i) from equipartition,
$\|P_1\psi_\gamma\|^2 = |A_1|$, for $|\Omega| \ll \Phi$ to localization on
$A_2$ for $|\Omega| \gg \Phi$ for various values of $\Phi/h$ and $|\Omega|/h$.
The transition is universal with the scaling parameter $|\Omega| / \Phi$.
Moreover, this even holds for individual states without averaging (red dots).
We stress that this localization transition in the open system occurs even
though $\Phi/h \geq 10$, where in the closed system all eigenstates are
equipartitioned~\cite{MicBaeKetStoTom2012}.

In Fig.~\ref{fig:fig2}, we fix the parameters such that $|\Omega| \gg \Phi$,
for which the long-lived resonance states localize on $A_2$, and show the
$\gamma$ dependence of the weights $\|P_1\psi_\gamma\|^2$ for all resonance
states. We find transition~(ii) from resonance states which localize on $A_2$
for small $\gamma$ to resonance states which localize on $A_1$ for large
$\gamma$, including equipartitioned resonance states in between.

The fact that both transitions (i) and (ii) occur for $h \ll \Phi$ suggests
that the localization transitions could be of classical origin. Furthermore,
from the point of view of decaying classical distributions the observed
transitions qualitatively seem to be rather intuitive: in Fig.~\ref{fig:fig1},
for a larger size of the opening one has less weight in region $A_1$. In
Fig.~\ref{fig:fig2}, a larger weight in $A_1$ corresponds to a larger decay
rate. For a quantitative description, however, one needs to find the quantum
mechanically relevant class of CIMs.

%%%%%%%%%%%%%%%%%%%%%%%%%%%%%%%%%%%%%%%%%%%%%%%%%%%%%%%%%%%%%%%%%%%%%%%%%%%%%%%

\emph{Classical localization.}---A conditionally invariant measure (CIM)
$\mu_\gamma$ is defined by
\begin{equation}\label{eq:c-measure_condition}
\mu_\gamma(T^{-1}(X)) = e^{-\gamma} \mu_\gamma(X),
\end{equation}
for each measurable subset $X$ of phase space. It is invariant under the
classical iterative dynamics $T$ of the open system up to an exponential decay
with rate $\gamma$. Equation~\eqref{eq:c-measure_condition} states that the
measure $\mu_\gamma(T^{-1}(X))$ of the set $T^{-1}(X)$ that will be mapped to
$X$ is smaller than $\mu_\gamma(X)$ by the factor $e^{-\gamma}$. These measures
must be zero on the iterates of the opening~$\Omega$. Thus, the support of
$\mu_\gamma$ is the fractal backward trapped set $\Gamma_\text{b}$ [horizontal
black stripes in Fig.~\ref{fig:fig3}(b)], that is the set of points in phase
space which do not escape under backward time evolution. Particularly important
is the natural CIM~$\mu_\text{nat}$, see Fig.~\ref{fig:fig3}(c), which is
constant on its support [because of integration over boxes in
Fig.~\ref{fig:fig3}(c) one finds two nonzero box measures].

We now generalize $\mu_\text{nat}$ to a CIM $\mu_\gamma$ of arbitrary decay
rate $\gamma$, which we call \emph{$\gamma$-natural CIM}. To this end, we use a
construction of CIMs~\cite{DemYou2006, NonRub2007} where one starts with an
arbitrary probability measure on the intersection $\Omega \cap \Gamma_\text{b}$
of the opening $\Omega$ with the backward trapped set $\Gamma_\text{b}$. By
propagating this measure backwards to all forward escaping sets
$T^{-n}(\Omega)$ [vertical colored stripes in Fig.~\ref{fig:fig3}(b)] and
appropriate scaling [respecting the decay rate $\gamma$,
Eq.~\eqref{eq:c-measure_condition}] one obtains a CIM. Here, we choose the
simplest measure on $\Omega \cap \Gamma_\text{b}$, given by $\mu_\text{nat}$.
This choice of a measure, which is constant on its support, is quantum
mechanically motivated in analogy to quantum ergodicity for closed fully
chaotic systems, where eigenstates in the semiclassical limit approach the
constant invariant measure~\cite{BaeSchSti1998, DegGra2003b}. This choice leads
to the $\gamma$-natural CIM
\begin{equation}\label{eq:cmeasure_arbitrary}
\mu_\gamma(X) = \mathcal{N} \, \sum_{n=0}^\infty
    e^{(\gamma_\text{nat} - \gamma)n} \mu_\text{nat}(X\cap T^{-n}(\Omega)),
\end{equation}
with normalization $\mathcal{N} = (1 - e^{-\gamma}) / (1 -
e^{-\gamma_\text{nat}})$. This series multiplies $\mu_\text{nat}$ in each
forward escaping set $T^{-n}(\Omega)$ by an appropriate factor which imposes
the overall decay rate $\gamma$ according to
Eq.~\eqref{eq:c-measure_condition}.
Two examples of $\gamma$-natural CIMs for the partial-barrier Baker map are
shown in Figs.~\ref{fig:fig3}(d) and \ref{fig:fig3}(e). The measure is constant
on $T^{-n}(\Omega)\cap \Gamma_\text{b}$ for each $n \in \mathbbm{N}_0$. With
increasing $n$, this constant is decreasing (increasing) for $\gamma >
\gamma_\text{nat}$ ($\gamma < \gamma_\text{nat}$); in particular, short-lived
measures $\mu_\gamma$ have more weight in the opening.
Note that the idea underlying Eq.~\eqref{eq:cmeasure_arbitrary} was used
without the notion of CIMs in Ref.~\cite{KeaNovPraSie2006} for sets $X =
T^{-n}(\Omega)$ for systems without a partial barrier.
Moreover, note that the $\gamma$-natural CIMs are solutions of the exact
Perron--Frobenius operator (which is not available), but cannot be obtained
from finite-dimensional approximations. Therefore, they have to be constructed
directly in phase space.

We find as our main result on the classical localization of $\mu_\gamma$ due to
a partial barrier that the weight of $\mu_\gamma$ on each side of the partial
barrier is given by~\footnote{See Supplemental Material}
\begin{equation}\label{eq:loc_classics_explicit}
\mu_\gamma(A_1) = \frac{\mu_\text{nat}(A_1) - c_\gamma}{1 - c_\gamma},
\end{equation}
and $\mu_\gamma(A_2) = 1 - \mu_\gamma(A_1)$, with
\begin{equation}\label{eq:gamma_coefficient}
c_\gamma = \left( 1 - e^{\gamma - \gamma_\text{nat}} \right) \,
    \left( 1 - e^{- \gamma_\text{nat}} \right) \,
    \frac{|A_1|}{|\Omega|} \frac{|A_2|}{\Phi}.
\end{equation}
The values for $\mu_\text{nat}(A_1)$ and $\gamma_\text{nat}$ follow from the
longest-lived eigenstate of the eigenvalue problem
\begin{equation}\label{eq:eval_problem_FPO}
F_\text{nat} \begin{pmatrix} \mu_\text{nat}(A_1) \\ \mu_\text{nat}(A_2)
\end{pmatrix}
    = e^{-\gamma_\text{nat}} \begin{pmatrix} \mu_\text{nat}(A_1) \\
        \mu_\text{nat}(A_2) \end{pmatrix},
\end{equation}
where $F_\text{nat}$ denotes the transition matrix between $A_1$ and $A_2$ for
the one-step propagation of $\mu_\text{nat}$ (see
Ref.~\cite{WebHaaBraManSeb2001} for approximations of the Perron--Frobenius
operator). In general, $F_\text{nat}$ may be obtained numerically or it may be
approximated by assuming a uniform distribution for $\mu_\text{nat}$,
\begin{equation}\label{eq:FPO_estimate}
F_\text{nat} \approx
\begin{pmatrix}
1 - (|\Omega| + \Phi) / |A_1| & \Phi / |A_2| \\
\Phi / |A_1| & 1 - \Phi / |A_2|
\end{pmatrix}.
\end{equation}
This turns out to be quite a good approximation even for fractal
$\mu_\text{nat}$ and it is exact for the partial-barrier Baker map.

%%%%%%%%%%%%%%%%%%%%%%%%%%%%%%%%%%%%%%%%%%%%%%%%%%%%%%%%%%%%%%%%%%%%%%%%%%%%%%%

\emph{Quantum-to-classical correspondence.}---Figure~\ref{fig:fig1} (green
line) shows the classical localization $\mu_\gamma(A_1)$,
Eq.~\eqref{eq:loc_classics_explicit}, for $\gamma = \gamma_\text{nat}$ (i.e.,
$c_\gamma = 0$ and $\mu_\gamma = \mu_\text{nat}$), using the approximation
Eq.~\eqref{eq:FPO_estimate}. When increasing the size $|\Omega|$ of the
opening, we find a transition from equipartition for $|\Omega| \ll \Phi$ to
localization for $|\Omega| \gg \Phi$. The only scaling parameters are
$|\Omega|/\Phi$ and $|A_1|/|A_2|$. We find very good agreement of the classical
localization measure with the localization of the quantum resonance states.
Note that for $\gamma\neq\gamma_\text{nat}$, the localization depends on all
parameters $|\Omega|$, $\Phi$, and $|A_1|$.

Figure~\ref{fig:fig2} (green line) shows Eq.~\eqref{eq:loc_classics_explicit}
as a function of $\gamma$. The classical localization measure $\mu_\gamma(A_1)$
monotonically increases with $\gamma$; i.e., the faster the decay, the larger
is the weight in region $A_1$ with the opening~$\Omega$. In the limit $\gamma
\to \infty$ one finds $\mu_\gamma(A_1) = 1$, and in fact, all the weight is in
the opening $\Omega$. In the limit $\gamma \to 0$ one finds a small constant
$\mu_0(A_1) > 0$; i.e., even though most of the weight is in $A_2$ there is
always a small contribution in $A_1$ due to the exchange between $A_1$ and
$A_2$.
Again, we find very good agreement between the classical localization measure
and the localization of the quantum resonance states for all decay rates
$\gamma$.
Note that quantum-to-classical correspondence is also confirmed for $|A_1| \neq
|A_2|$ (not shown).

%%%%%%%%%%%%%%%%%%%%%%%%%%%%%%%%%%%%%%%%%%%%%%%%%%%%%%%%%%%%%%%%%%%%%%%%%%%%%%%

\begin{figure}
\begin{center}
\includegraphics{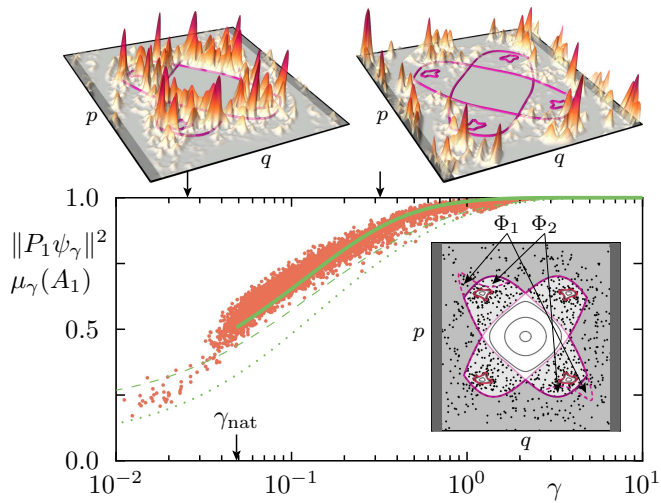}
\caption{
(color online) Weight $\|P_1\psi_\gamma\|^2$ (red points) of resonance states
$\psi_\gamma$ in region $A_1$ vs decay rate $\gamma$ for standard map at
$\kappa = 2.9$, with $|A_1| \approx 0.6664$, $|A_2| \approx 0.2061$, $\Phi
\approx 0.0126$, $|\Omega| = 0.1$, and $h=1/10000$. This is compared to the
$\gamma$-natural CIM $\mu_\gamma(A_1)$, either by direct numerical
computation~\cite{Note1} (solid green line), using
Eq.~\eqref{eq:loc_classics_explicit} and computing $F_\text{nat}$ numerically
(dashed green line), or by using approximation Eq.~\eqref{eq:FPO_estimate}
(dotted green line).
Inset: Phase space of the standard map with regular and chaotic regions,
illustrating regions $A_1$ (medium gray shaded), $A_2$ (light gray shaded) on
either side of the main partial barrier (thick solid magenta line) with
exchanging regions $\Phi_1$, $\Phi_2$, and opening $\Omega$ (dark gray shaded).
Upper panels: Husimi representation of typical long-lived (left) and
short-lived (right) resonance states for $h=1/1000$ with $\gamma$ values
indicated by arrows.
}
\label{fig:fig4}
\end{center}
\end{figure}
Do these results for the partial-barrier map generalize to generic systems? In
Fig.~\ref{fig:fig4}, we show for the standard map at $\kappa = 2.9$, where it
has a mixed phase space, that the localization of the chaotic resonance states
on region $A_1$, which contains the opening, increases as a function of
$\gamma$. Qualitatively, we find the same localization behavior as for the
partial-barrier standard map in Fig.~\ref{fig:fig2}. Quantitatively, it is well
described by the classical localization of $\mu_\gamma$, which is determined
numerically~\cite{Note1}. Also the analytical prediction,
Eq.~\eqref{eq:loc_classics_explicit}, works reasonably well. Overall,
Figs.~\ref{fig:fig1}, \ref{fig:fig2}, and \ref{fig:fig4} demonstrate
quantum-to-classical correspondence for the localization of chaotic resonance
states in open systems due to a partial barrier.

%%%%%%%%%%%%%%%%%%%%%%%%%%%%%%%%%%%%%%%%%%%%%%%%%%%%%%%%%%%%%%%%%%%%%%%%%%%%%%%

\emph{Outlook.}---We see the following future challenges: (a)~While in this
work we concentrate on the weights on either side of a partial barrier one
should verify the quantum-to-classical correspondence for the fine-structure of
chaotic resonance states to $\gamma$-natural CIMs. (b)~Which deviations arise
when approaching the quantum regime of $h \approx \Phi$, $|\Omega|$? (c)~Is the
new class of $\gamma$-natural CIMs, which is quantum mechanically motivated, of
relevance also in classical dynamical systems? (d)~Is it possible to predict
which quantum mechanical decay rates $\gamma$ occur in the presence of a
partial barrier including their distribution, as it is known for fully chaotic
systems~\cite{ZycSom2000, NonZwo2009, MicAlt2013}? (e)~The present work
explains the localization of resonance states which have been used to derive
the hierarchical fractal Weyl laws~\cite{KoeMicBaeKet2013} for a hierarchy of
partial barriers. Now it is possible to discuss whether these laws survive in
the semiclassical limit. (f)~We see direct applications to mode coupling in
optical microcavities~\cite{Wie2014b} and in recently studied
parity--time symmetric systems~\cite{WesKotPro2010, Sch2013b}, where instead of
a partial barrier one has coupled symmetry-related subspaces.

\begin{acknowledgments}
We are grateful to E.\,G.~Altmann, K.~Clau{\ss}, S.~Nonnenmacher, and
H.~Schomerus for helpful comments and stimulating discussions, and acknowledge
financial support through the Deutsche Forschungsgemeinschaft under Grant
\linebreak No.\ KE 537/5-1.
\end{acknowledgments}


%%%%%%%%%%%%%%%%%%%%%%%%%%%%%%%%%%%%%%%%%%%%%%%%%%%%%%%%%%%%%%%%%%%%%%%%%%%%%%%

\section{Supplemental Material}

\emph{Classical derivation.}---In order to derive
Eq.~\eqref{eq:loc_classics_explicit}, we focus on the localization properties
of $\mu_\gamma$ with respect to the partial barrier, restricting ourselves to
the partial-barrier Baker map in the following. The generalization to other
systems will be discussed at the end.

The localization of $\mu_\gamma$ is described by its weights $\mu_\gamma(A_k)$
on either side of the partial barrier. In virtue of Eq.~(3), we only have to
study $\mu_\text{nat}(A_k\cap T^{-n}(\Omega))$ in more detail to compute
$\mu_\gamma(A_k)$. We find that the natural measure of $A_k\cap T^{-n}(\Omega)$
is proportional to its relative area inside $A_k$,
\begin{equation}\label{eq:productstructure}
\mu_\text{nat}(A_k\cap T^{-n}(\Omega))
= \mu_\text{nat}(A_k) \cdot \frac{|A_k \cap T^{-n}(\Omega)|}{|A_k|}. \tag{S1}
\end{equation}
This follows from the fact that the forward escaping sets $T^{-n}(\Omega)$
decompose the backward trapped set $\Gamma_\text{b}$ in the unstable
(horizontal) direction, on which $\mu_\text{nat}$ is uniformly distributed
within $A_1$ and $A_2$ individually, see Fig.~3(c).

The distribution of the opening~$\Omega$ over phase space under backward time
evolution, which enters Eq.~\eqref{eq:productstructure} in terms of $|A_k \cap
T^{-n}(\Omega)|$, follows from
\begin{equation}\label{eq:op_leftright_FPO}
\begin{pmatrix} |A_1 \cap T^{-n}(\Omega)| \\ |A_2 \cap T^{-n}(\Omega)|
\end{pmatrix}
= F_\text{nat}^n \begin{pmatrix} |\Omega| \\ 0 \end{pmatrix} , \tag{S2}
\end{equation}
where $F_\text{nat}$ denotes the transition matrix between $A_1$ and $A_2$ for
the one-step propagation of $\mu_\text{nat}$. Note that the transition matrix
for the backward time evolution of $\Omega$ is given by $F_\text{nat}$ itself.
This relation can be interpreted using Fig.~3(b): In the beginning, $\Omega$
(gray vertical stripe) is supported on $A_1$. In the next step,
$T^{-1}(\Omega)$ (yellow) splits into equal parts on $A_1$ and $A_2$.
Afterwards, $T^{-2}(\Omega)$ (orange) contributes two stripes to $A_1$ and
three to $A_2$.

Inserting the relations~\eqref{eq:productstructure} and
\eqref{eq:op_leftright_FPO} in Eq.~(3), and using Neumann's series, we obtain
\begin{equation}\label{eq:loc_classics}
\mu_\gamma(A_k) = \mathcal{N} \frac{\mu_\text{nat}(A_k)}{|A_k|}
    \left[ \bigl(\mathbbm{1} - e^{\gamma_\text{nat} - \gamma}
F_\text{nat}\bigr)^{-1}
    \begin{pmatrix} |\Omega| \\ 0 \end{pmatrix} \right]_k . \tag{S3}
\end{equation}
Our result on the classical localization, Eq.~(4), follows from
Eq.~\eqref{eq:loc_classics} after some algebra.

We believe that this derivation for the partial-barrier Baker map can be
generalized to generic dynamical systems. This is indicated by the numerical
findings for the partial-barrier standard map, Fig.~2, as well as the standard
map, Fig.~4, compared to Eq.~(4). The main reason for the deviations of Eq.~(4)
for a generic system is that Eq.~\eqref{eq:productstructure} is not valid in
general. For its generalization, one has to revise the argument in the natural
coordinates of the stable and unstable manifolds.

\emph{Standard map.}---In the following, we explain how to determine
$\mu_\gamma$ for a generic system like the standard map numerically. First, one
has to approximate (the chaotic part of) the backward trapped
set~$\Gamma_\text{b}$. To this end, one may define a uniform grid of
$N_\text{grid}$ points in phase space of which one has to discard points which
leave the system within $N_\text{iter}$ iterations of the map~$T$ in backward
time direction. For Fig.~4, we choose $N_\text{grid} = 10^6$ and $N_\text{iter}
= 50$. Points within a regular phase-space region should be omitted manually.
The remaining points provide the finite-time
approximation~$\Gamma_\text{b}^\text{num}$ of $\Gamma_\text{b}$ and need to be
classified by their forward escaping times. Following Ref.~[18], we associate
the $\mu_\gamma$ measure $e^{-\gamma n}(1 - e^{-\gamma})$ to set
$T^{-n}(\Omega) \cap \Gamma_\text{b}^\text{num}$. Finally, assuming
equidistribution for the points in $T^{-n}(\Omega) \cap
\Gamma_\text{b}^\text{num}$, we find
\begin{equation}
\mu_\gamma(X\cap T^{-n}(\Omega)) \approx f_n(X) \, e^{-\gamma n}(1 -
e^{-\gamma}),
\tag{S4}
\end{equation}
for each (measurable) subset $X$ of phase space and
\begin{equation}
f_n(X) := \frac{\# \left(X \cap T^{-n}(\Omega) \cap \Gamma_\text{b}^\text{num}
\right)} {\#  \left(  T^{-n}(\Omega) \cap \Gamma_\text{b}^\text{num} \right)}
\tag{S5}
\end{equation}
where $\#$ denotes the cardinality of a set. Using $\mu_\gamma(X) =
\sum_{n=0}^{\infty} \mu_\gamma(X\cap T^{-n}(\Omega))$ we have a numerical
estimate for the $\gamma$-natural CIM $\mu_\gamma$. As the sample
$\Gamma_\text{b}^\text{num}$ is only finite the series will terminate and the
numerically approximated measure is not perfectly normalized. This method is
not appropriate for exceedingly small $\gamma$ since the weight on forward
escaping sets $T^{-n}(\Omega)\cap\Gamma_\text{b}^\text{num}$ with large escape
times $n$ increases while they are approximated by a few points only. Hence, we
restrict the method to $\gamma \geq \gamma_\text{nat}$ in Fig.~4.

Quantum mechanically, owing to the mixed phase space of a generic system, we
discard all regular and deeper hierarchical states having less than 50\% of
their weight within $A_1$ and $A_2$. As some of the remaining chaotic resonance
states still have a significant contribution outside of $A_1 \cup A_2$, we
renormalize them such that $\|P_1 \psi_\gamma\|^2 + \|P_2 \psi_\gamma\|^2 = 1$.

\end{document}